\documentclass[lettersize,journal]{IEEEtran}
\usepackage{amsmath,amsfonts}
\usepackage{algorithmic}
\usepackage{algorithm}
\usepackage{array}
\usepackage{textcomp}
\usepackage{stfloats}
\usepackage{url}
\usepackage{verbatim}
\usepackage{subcaption}
\usepackage{cite}
\usepackage{graphicx}
\usepackage{longtable}
\usepackage{lscape}

\newcommand{\etal}{\textit{et al}. }

\begin{document}

\title{Machine Learning Approaches for Fine-Grained Symptom Estimation in Schizophrenia: \\ A Comprehensive Review \\

}

\author{\IEEEauthorblockN{Niki Maria Foteinopoulou and Ioannis Patras} \\
\IEEEauthorblockA{\textit{School of Electronic Engineering and Computer Science} \\
\textit{Queen Mary University of London}\\
London, United Kingdom \\
\{n.m.foteinopoulou, i.patras\}@qmul.ac.uk}

}




\maketitle

\begin{abstract}
Schizophrenia is a severe yet treatable mental disorder, whose definition has evolved significantly since its inception in the early 20th century. Initially conceived as a broad term encompassing various serious mental health conditions, it is being diagnosed using a multitude of primary and secondary symptoms. Diagnosis and treatment for each individual depends on the severity of the symptoms, therefore there is a need for accurate, personalised assessments. However, while diagnostic and assessment standards exist, the process can be both time-consuming and subjective; hence, there is a compelling motivation to explore automated methods that can offer consistent diagnosis and precise symptom assessments, thereby complementing the work of healthcare practitioners. Machine Learning, a dominant paradigm in Artificial Intelligence, has demonstrated impressive capabilities across numerous domains, including medicine. The use of Machine Learning in patient assessment holds great promise for healthcare professionals and patients alike, as it can lead to more consistent and accurate symptom estimation.
This survey paper aims to review methodologies that utilise Machine Learning for diagnosis and assessment of schizophrenia.  Contrary to previous reviews that primarily focused on binary classifications distinguishing patients from healthy control groups, this work recognises that schizophrenia is a complex condition with manifestations that extend beyond a simple binary categorisation and instead, offers an overview of Machine Learning methods designed for fine-grained estimation of schizophrenia symptoms. We cover multiple modalities, namely Medical Imaging in the form of Magnetic Resonance Imaging, Electroencephalograms and Audio-Visual input, as the illness symptoms can manifest themselves both in a patient's pathology and behaviour. Finally, we analyse the machine learning methodologies used in the studies included in the survey and identify trends and gaps in the literature and opportunities for future research. 
\end{abstract}

\begin{IEEEkeywords}
Fine-grained labels, Schizophrenia, Mental Health, Machine Learning
\end{IEEEkeywords}

\section{Introduction}

 Schizophrenia is a mental disorder with debilitating effects~\cite{saha_systematic_2005, moreno-kustner_prevalence_2018}; the term schizophrenia, first appeared by Eugen Bleuler in 1908, in an attempt to redefine what until that point was thought to be premature dementia~\cite{bleuler1908prognose}. At the time, the condition was thought to be a separation in personality, thinking, and general cognitive function, as described by the components of the term which translate from ancient Greek to \textit{to split} and \textit{mind}. Historically, there has been a great misunderstanding of the condition by both the general population and early psychiatrists, often used as a blanket diagnosis for very serious mental illnesses. As research progressed, the understanding of the illness has been improved and the definition has been narrowed down. According to the Diagnostic and Statistical Manual of Mental Disorders, Fifth Edition (DSM-V)~\cite{dsmd_v}, for a diagnosis of schizophrenia the patient needs to demonstrate at least two symptoms of the primary categories after at least one episode of psychosis. More specifically, one of the symptoms needs to be hallucinations, delusions or disorganised speech and a second symptom can be one of the negative symptoms~\cite{kay_positive_1987} (eg. Blunted Affect). However, post-diagnosis, and similarly to most mental illnesses, several secondary symptoms are associated with the disease which makes each diagnosis unique and the illness diverse overall. The complete spectrum of primary and secondary symptoms, along with their respective intensities affect the treatment course and are therefore as important as the primary diagnosis.

 In recent years, Artificial Intelligence (AI) and more specifically Machine Learning (ML) have dominated the news and public discussions. ML refers to the field of AI, that leverages statistical techniques to learn patterns from data and has become the dominant paradigm in fields such as Computer Vision~\cite{dalle} and Natural Language Processing~\cite{gpt2}. As the capabilities of such technology increase and boundaries are pushed, the discussion of how such technology can improve daily life is becoming more relevant than ever. In mental health diagnosis and assessment, and more specifically for schizophrenia, automated methods would greatly assist professionals and patients by offering consistent and accurate diagnoses.
 In clinical practice, the DSM-V~\cite{dsmd_v} provides a standardised framework for conducting interviews, along with guidelines for scoring symptoms on various scales~\cite{kay_positive_1987, forbes2010initial}. However, practitioners face limitations in directly observing and quantifying patient behaviour during the interview process. Instead, they often rely on post-interview assessments based on overall patient behaviour, self-reports, and input from the patient's family. Having a second practitioner who would be directly observing and quantifying verbal and non-verbal cues during the interview would offer greater accuracy in symptom assessment. However, such an approach would be significantly more time-consuming and labour-intensive, according to some estimations taking over ten times longer~\cite{troisi_ethological_1999}.

 Contrary to other medical fields where several methodologies that aim to assist with diagnosis have been developed~\cite{ma2023segment, he2023accuracy, bransby20233d}, mental health remains relatively unexplored; this can be attributed to several problems associated with the problem formulation but also the data availability in the field. More specifically, mental health problems often rely on self-reporting~\cite{avec2013}; when self-reporting is not an option, expert annotations are needed~\cite{ness}, which can be costly to obtain, particularly for fine-grained symptom level annotations. Finally, as for the assessment of several illnesses and disorders non-verbal cues need to be assessed (e.g., facial expressions); these are private data, and there are few available datasets to researchers which typically contain few samples.

 Existing surveys on automated methods focus on diagnosis with a binary decision for schizophrenia~\cite{cortes-briones_going_2022}; this is the case as binary classification between patients and healthy controls is the dominant paradigm with few works using ML identify sub-categories and, more specifically, individual symptoms. Furthermore, previous surveys often focus on single modalities such as computer vision or speech and language~\cite{jiang_utilizing_2022, voleti_review_2020}, thus disregarding findings and patterns by cross-sectionally analysing literature in fine-grained schizophrenia assessment.
 More specifically, these are focusing on three main research streams, namely medical imaging with the use of structural or functional Magnetic Resonance Imaging (MRI)~\cite{swati_machine_2020}, bio-signals in the form of electroencephalogram (EEG)~\cite{barros_advanced_2021}, and audio-visual input~\cite{jiang_utilizing_2022, voleti_review_2020}. 
 While in each modality there are different aspects of the illness examined, by comparing works across streams we can uncover common challenges faced by researchers as well as complementary information regarding the illness. For example, a large aspect of medical imaging and EEG signals is making an assessment using pathological and anatomical effects of the illness on the brain, while also uncovering the underlying dysregulation in the brain structure or function~\cite{oh_identifying_2020, chu_individual_2018}. On the other hand, several works using audio-visual input explicitly or implicitly study patient behaviour~\cite{abbas_computer_2021, vail_visual_2017}. However, in both researchers are faced with limited data availability due to a lack of fine-grained annotations~\cite{mridataset, fmri1, fmri2} or confidentiality constraints since facial and vocal information are private data. In addition, while input types can greatly affect the methodology used, there are also parallels and common techniques, for example, the use of Convolutional Neural Networks (CNN) across imaging sources (medical or RGB input)~\cite{qureshi_3d-cnn_2019, oh_deep_2019, bishay_schinet_2021, foteinopoulou_learning_2022}.
 Furthermore, we review the datasets used in the studies identified in this survey and compare the data collection, annotation methods and availability, which is a novel contribution of this survey.

 The contributions of this survey can be summarised as follows:
 \begin{itemize}
     \item This survey is the first of its kind, focusing on fine-grained assessment of patients with schizophrenia. Unlike previous surveys that focus explicitly~\cite{cortes-briones_going_2022} or implicitly~\cite{jiang_utilizing_2022} on binary classification between schizophrenia patients and healthy controls, we focus on works that identify sub-categories or estimate symptom severity and are thus more reflective of real-world diagnostic conditions.
     \item This survey analyses cross-sectionally the methods used across multiple modalities, namely MRI, EEG and Audio-Visual. By concurrently reviewing the works across all modalities, we can uncover common patterns in the ML methods used and challenges faced by researchers.
     \item We compare the collection, annotation and availability of datasets used in the works included in this survey, which is not addressed by previous surveys on ML methods in schizophrenia diagnosis and assessment~\cite{cortes-briones_going_2022, jiang_utilizing_2022, voleti_review_2020, swati_machine_2020, barros_advanced_2021}.
     \item We conduct a comprehensive discussion on the approaches, challenges and gaps in current research. In addition, we discuss the most promising research directions in the fine-grained schizophrenia assessment task.
 \end{itemize}

 The remainder of the paper is organised as follows. First, we discuss the background and some key concepts associated with schizophrenia in Section~\ref{sec:background}. We outline the search strategy to find and identify relevant works in Section~\ref{sec:method}. We then review the works identified in the literature, describing the ML techniques in Section~\ref{sec:review}. Finally, we discuss the current research in employing ML in mental health, along with its open questions in Section~\ref{sec:discussion}.

\section{Background and Preliminary Concepts}
\label{sec:background}

\begin{figure}
    \centering
    \includegraphics[width=\linewidth]{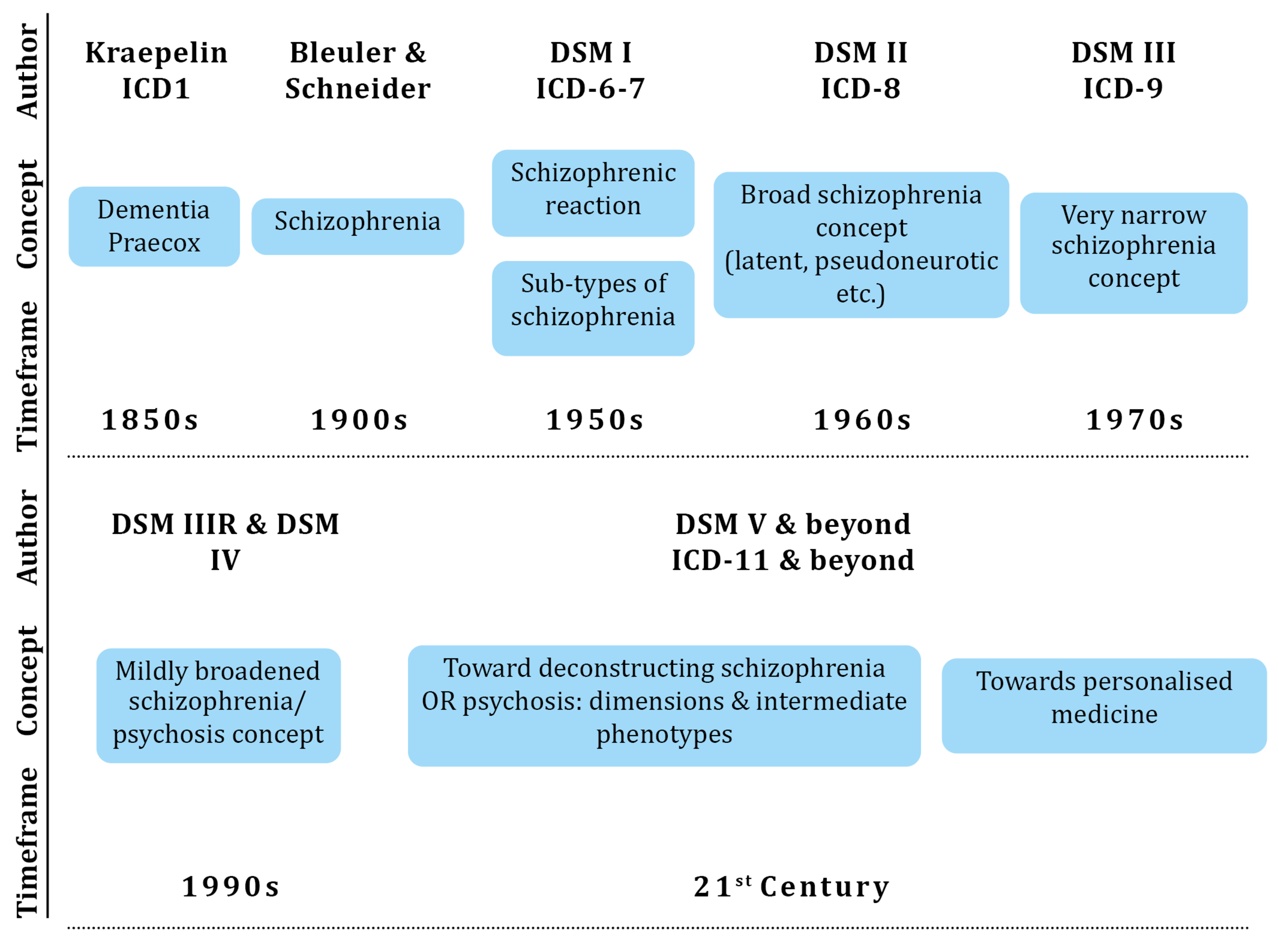}
    \caption{A simplified timeline of the illness definitions from the 1850s to now.}
    \label{fig:timeline}
\end{figure}
 Schizophrenia is a mental illness that while it has existed for millennia, with its symptoms described in early literature; however, its identification as a mental illness, is more recent tracing back to the late 19th century~\cite{tamminga2005phenotype}. 
 Eugen Bleuler first introduced the definition of schizophrenia in 1908~\cite{bleuler1908prognose}, however, the understanding of the illness has undergone several revisions over the years, with manuals and guidelines also updated (Fig.~\ref{fig:timeline}). Even though schizophrenia is a serious mental illness, several misconceptions and stigmas have been associated with the disease in popular belief~\cite{messias2007epidemiology}. To establish a solid understanding of the problem and how machine learning tools can be used, in this section, we briefly discuss the nature of the illness, the current diagnostic tools physicians use, and how affective computing can assist in improving mental health diagnosis and treatment.

\subsection{Diagnosis of Schizophrenia}
 For the diagnosis of mental health illnesses, including schizophrenia spectrum disorder commonly referred to as schizophrenia, most practitioners follow the diagnostic criteria set in either DSM-V~\cite{dsmd_v} or International Statistical Classification of Diseases and Related Health Problems (ICD)~\cite{icd10}. While some definitions and categorisations have changed between editions, the diagnostic criteria are largely the same between the two manuals; for a diagnosis of schizophrenia, patients need to demonstrate at least two symptoms of the illness and some social dysfunction as a result of the disease for a prolonged period.

 More specifically, at least one core symptom (i.e., delusions, hallucinations, or disorganised speech/thought) needs to be present as well as at least one additional symptom (i.e., disorganised or catatonic behaviour, negative symptoms, or cognitive symptoms). Symptoms should be present for most of the time during at least one month and significantly affect the level of functioning in at least one area (such as work, interpersonal relations, etc.). 

 As there is great variability in the diagnostic criteria that may require both observation and reports of social behaviour~\cite{kay_positive_1987, dsmd_v, icd10}, there is no single test that can be performed to diagnose the illness. Furthermore, as several of the symptoms of schizophrenia are also symptoms of other conditions~\cite{joyce1984age, lisanby1998psychosis, koksal2015case, schultz2007schizophrenia}, physicians need to perform several physical and psychiatric evaluations to rule out symptoms due to substance abuse, medication or other conditions. Overall, both due to the diagnostic criteria and the time required to confirm a diagnosis (including any logistical and systemic delays), the time to reach a definitive diagnosis can vary from weeks to months.

\subsection{Secondary Symptoms, Pathology and Treatment}
 Post-diagnosis, several secondary symptoms are associated with the illness. Several symptom severity scales have been proposed over the years~\cite{kay_positive_1987, forbes2010initial, nsa16}, focusing on different aspects, namely: positive, negative, and general psychopathology. The category names refer to the presence of symptoms in patients relative to the general population; therefore, positive symptoms such as hallucinations or delusions, refer to symptoms and behaviour that are present in schizophrenia patients but not in the general population, negative symptoms such as blunted affect describe behaviours that are absent in patients of schizophrenia, and finally general psychopathology includes other symptoms such as depression or anxiety. However, while symptoms and severity scales are well defined, there is often disagreement among experts when assessing patients~\cite{shrout1998measurement}.

 Further to the behavioural symptoms, several studies have identified areas of the brain associated with the illness~\cite{zeng_multi-site_2018, oh_identifying_2020, li_machine_2019, brown2010functional}. Although using the patient's brain pathology is not part of the routine diagnostic process, brain areas associated with pathological features may be targeted by pharmacological interventions and therapy in the future.


\section{Methods for Literature Selection}
\label{sec:method}

\subsection{Search Methodology}
 For the purpose of this review, we employed a scoping review methodology, which involves mapping the key concepts that underpin the research area~\cite{arksey2005scoping}, as opposed to adopting a systematic approach that would draw evidence from a more limited number of studies.
    \begin{figure}[t]
            \includegraphics[width=\linewidth]{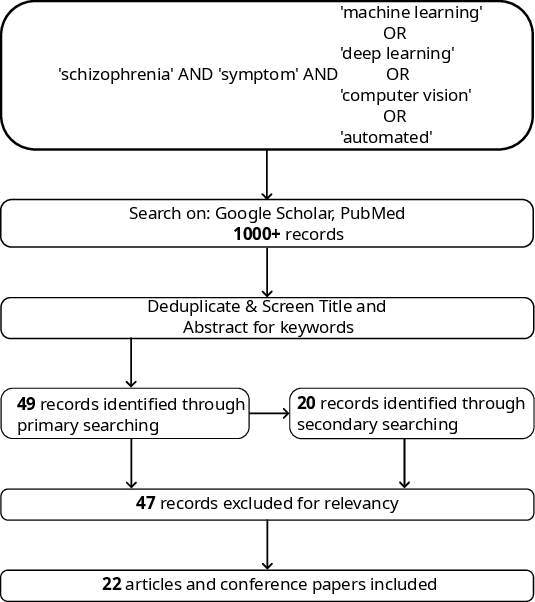}
            \caption{Flow diagram for literature selection.}
            \label{fig:flow}
    \end{figure}

Our search strategy has two stages, illustrated in Fig.~\ref{fig:flow}. The initial stage encompasses a primary search that involves querying various databases. In this stage, our focus lies in conducting a comprehensive search across the querying databases using keywords. Subsequently, the second stage involves a supplementary search within the references cited by the studies that were identified during the primary search. The intention behind this secondary stage is to uncover additional relevant works that might have been overlooked in the initial search. 

More specifically, the primary search involved querying Google Scholar~\cite{scholar} and PubMed~\cite{pubmed}, with a filter to include works published between January 2014 and August 2023, for articles and conference papers using a range of keywords, including ``schizophrenia'', ``symptom'', ``machine learning'', ``deep learning'', ``computer vision'', and ``automated''. A secondary search to include any relevant articles that are cited by articles in the primary search but are not included in the search results was conducted by screening the titles and abstracts of the cited works for relevancy.  

\subsection{Inclusion and Exclusion Criteria}
 Following these search phases, we conducted a thorough evaluation of the compiled literature to assess its relevance to our research objectives. We selected twenty-two articles for inclusion in our review from this screening process.
 The inclusion criteria for the papers retrieved were the following:
\begin{itemize}
    \item The methodology includes some machine learning/automated techniques.
    \item The method is evaluated on at least 10 patients, to have more degrees of freedom and thus statistically significant results.
    \item The method is evaluated on sub-categories or individual symptoms of schizophrenia (fine-grained approach).
\end{itemize}

On the contrary, articles that fall into at least one of the following were excluded:
\begin{itemize}
    \item The method addresses the problem only as a binary classification between healthy control and schizophrenia patients.
    \item The work has not been through a peer-review process (i.e. is only available on pre-print)
\end{itemize}


\section{Survey of Machine Learning Methods in Fine-grained Symptom Severity Estimation in Schizophrenia}
\label{sec:review}

 Numerous ML techniques have been proposed to address the estimation of symptom severity in schizophrenia. These approaches can be broadly categorised into three main work-streams: 1) \textbf{Medical Imaging}, primarily utilising fMRI data, 2) \textbf{EEG signal processing}, and 3) \textbf{Behavioural Analysis}, which involves analysing audio-visual footage of clinical interviews.

 These three streams align with either the pathological or behavioural manifestations of the illness, as discussed in Section~\ref{sec:background}. Specifically, studies have revealed that the non-verbal behaviour of individuals with schizophrenia undergoes changes corresponding to symptom severity~\cite{dimic2010non, lavelle2013nonverbal, worswick2018negative}. For instance, individuals scoring high on the negative scale tend to exhibit reduced eye contact and diminished smiling~\cite{dimic2010non}. Similarly, previous research extensively reviewed and documented differences in brain structure and function among patients with schizophrenia~\cite{antonova2004relationship, itil1977qualitative}. Similarly, MRI studies have consistently reported a volume reduction in brain areas of schizophrenia patients~\cite{tamminga2005phenotype}.
Therefore, by explicitly (i.e. hand-crafted features) or implicitly (from raw image or signal) using pathological or behavioural features as input, ML methodologies can learn the underlying pattern of the illness.

    \subsection{Medical Imaging}

    Within the domain of diagnosis and symptom estimation of schizophrenia using medical imaging, two types of MRI techniques are commonly employed: structural MRI and functional MRI (fMRI). The utilisation of these techniques offers distinct advantages and enables researchers, as well as automated systems, to gain valuable insights into the underlying characteristics of the brain.

    Structural MRI, as the name suggests, focuses on capturing detailed images of the brain's structure. It provides high-resolution visualisations of anatomical features, allowing human experts and automated systems to detect and analyse differences in brain morphology. By examining the structure of various brain regions, researchers can identify potential abnormalities or deviations that may be indicative of schizophrenia~\cite{shenton2001review, haukvik2013schizophrenia}.
    
    On the other hand, functional MRI (fMRI) operates on a different principle, measuring changes in blood flow within the brain. This technique relies on the observation that alterations in neural activity are typically accompanied by corresponding changes in local blood supply. By monitoring blood oxygenation levels, fMRI can map brain regions that are activated during specific tasks or in resting states. This enables researchers to investigate functional connectivity patterns and identify neural networks associated with schizophrenia-related symptoms~\cite{mitchell2001fmri, gur2002fmri}.
    
\begin{figure}
      \centering
      \begin{tabular}{@{}c@{}c@{}}
        \includegraphics[width=0.5\linewidth]{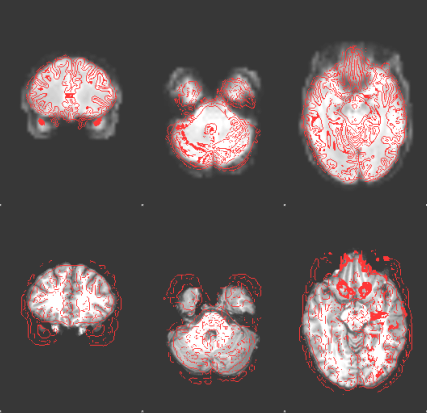} 
        
        &

        \includegraphics[width=0.5\linewidth]{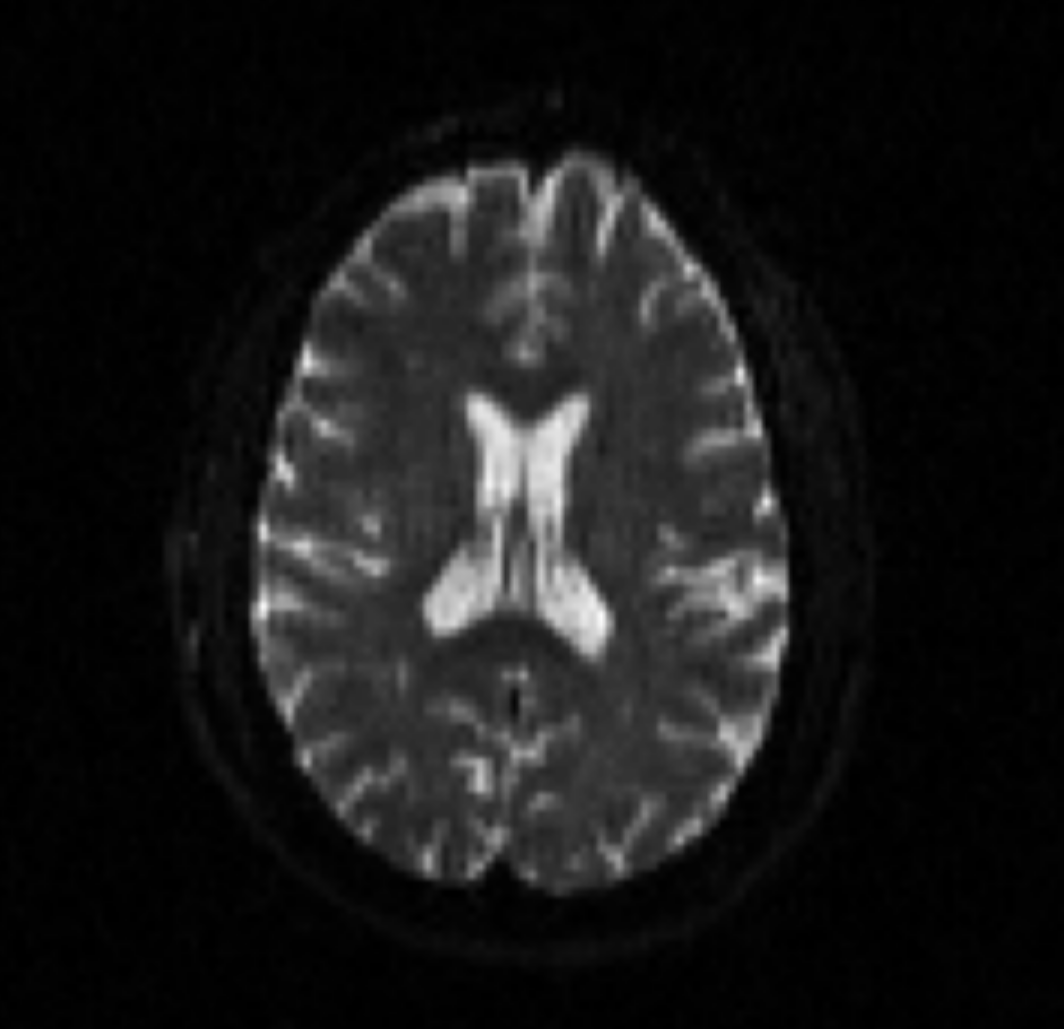} \\[\abovecaptionskip]
        
        \small (a) fMRI & \small (b) structural MRI
      \end{tabular}
    
      \caption{Examples of functional MRI (a), and structural MRI (b) from OpenfMRI~\cite{fmri1, fmri2} and SchizConnect~\cite{SchizConnect} respectively.}
    \label{fig:mri}
    \end{figure}

    Examples of MRI and fMRI can be seen in Fig.~\ref{fig:mri}, from two publicly available MRI datasets, namely the OpenfMRI~\cite{fmri1, fmri2} and the MCIC collection~\cite{mridataset} which are available through SchizConnect~\cite{SchizConnect}.

    Several studies use structural MRIs as the input to diagnose schizophrenia in a binary manner~\cite{talpalaru_identifying_2019,zheng_diagnosis_2021, oh_identifying_2020, swati_machine_2020, sadeghi_overview_2022}; as a matter of fact, the use of MRIs for binary schizophrenia classification is the dominant paradigm~\cite{cortes-briones_going_2022}. However, our emphasis lies on works that undertake a more nuanced approach to diagnosis, such as coarse sub-categories of schizophrenia or symptom estimation. This refined focus restricts the quantity of relevant studies available.

    Talpalaru~\etal~\cite{talpalaru_identifying_2019}, extracted hand-crafted features from structural MRIs of 167 subjects to train and compare a Logistic Regression, a Support Vector Machine (SVM), and a Random Forest algorithms in multi-class classification for symptom presence and intensity. Specifically, as the symptoms of schizophrenia are not mutually exclusive to each other and can occur concurrently, they used hierarchical clustering to derive subgroups from the symptoms and their associated intensities. More specifically, using Agglomerative hierarchical clustering three sub-groups of schizophrenia patients are identified, namely: a) high symptom burden, b) predominantly positive, c) low symptom burden and d) healthy control. The method is evaluated using Area Under the ROC Curve (AUC) metric, achieving an AUC of up to 81\% with better results on the high-symptom class. Moreover, the linear nature of the chosen algorithms provides a degree of interpretability to the results; the authors, use the feature importance of Random Forest to evaluate the relationship between regional impairments in the brain and symptoms of schizophrenia. However, such an approach requires expert knowledge for feature extraction from the MRI. Furthermore, the four classes obtained through the symptom clustering are a clear improvement from binary classification but do not adequately represent the spectrum of the disease, as the presence and severity of negative symptoms are excluded.

    Using SVM and hand-crafted features from MRIs, Gould~\etal~\cite{gould2014multivariate} performed a three-class classification between healthy participants, ``cognitive deficit'' patients, and ``cognitive spared'' patients identified in previous studies~\cite{green2013genome, jablensky2006subtyping, hallmayer2005genetic}. The method attains a relatively high accuracy, reaching up to 72\% with statistically significant results when compared to chance which is 63\% for the majority class "cognitive spared" in the selected dataset. These sub-types focus on the cognitive aspect of the illness which is a secondary symptom and often not the main focus, but do not consider any other primary or secondary symptoms. 

    Chand~\etal~\cite{chand2020two} proposed a self-supervised approach using HYDRA~\cite{varol2017hydra}, a method that performs SVM classification and clustering simultaneously to identify sub-types within the patient group. Contrary to traditional k-means clustering methods, the approach effectively clusters patients based on their differences from controls. The sub-types identified showed significant differences in brain anatomy which could pave the road for more personalised pharmaceutical treatment. However, there is no analysis of how symptoms manifest to the identified sub-types. Similarly, Honnorat~\etal~\cite{honnorat2019neuroanatomical} use a semi-supervised method to identify sub-types of patients based on fMRI scans. The method also tests for statistical significance in the PANSS~\cite{kay_positive_1987} scores between the identified groups, however, no statistically significant difference in the symptom severity exists between the identified sub-groups.

    The literature using fMRI for diagnosis and symptom estimation follows a similar theme to that of structural MRIs, primarily using hand-crafted features and linear or shallow methodologies. Using voxels from fMRIs, Bleich-Cohen~\etal~\cite{bleich-cohen_machine_2014} trained an SVM classifier to perform a three-class classification between healthy controls and schizophrenia patients with and without Obsessive–Compulsive Disorder (OCD). The method achieved accuracy up to 91\% with statistically significant results. Similarly, Chyzhyk~\etal~\cite{chyzhyk2015discrimination}, used hand-crafted features from fMRI to train an SVM on a three-class classification task. Specifically, the authors trained classifiers to distinguish between healthy controls, patients with auditory hallucinations (one of the most common symptoms of schizophrenia), and patients without auditory hallucinations. The method achieved very high accuracy, over 95\% with statistically significant results. This study also examined the areas of the brain that contribute to auditory hallucinations. Previous fMRI studies have addressed the problem in a binary manner, classifying samples into patient or control groups as previously discussed; while Chyzhyk~\etal~\cite{chyzhyk2015discrimination} took a significantly more fine-grained approach, the study focused on a single symptom, thus not exploring the full illness spectrum.
    
    The most fine-grained approach to symptom estimation using medical imaging identified in this survey is the method proposed by Tolmeijer~\etal~\cite{tolmeijer_using_2018}. The authors used fMRI to measure the effect of Cognitive Behavioural Therapy (CBT) on positive and depressive symptoms on the PANSS symptom scale~\cite{kay_positive_1987}. Using hand-crafted features and multivariate regression, Tolmeijer~\etal~\cite{tolmeijer_using_2018} estimated the improvement of each symptom from CBT. The method reported a Pearson's Correlation Coefficient (PCC) of 63\% for the positive psychotic symptoms and 31\% for depressive symptoms. 
    While this work addresses several symptoms rather than coarse sub-groups, it still relies on hand-crafted features. Furthermore, the study approached the problem in a relatively simple manner which is a strength, however, as subsequent works have shown in similar tasks~\cite{dakka_learning_2017,qureshi_3d-cnn_2019,zheng_diagnosis_2021, zeng_multi-site_2018}, there are improvements in terms of the evaluation metrics with the use of deep learning methodologies.

    A summary of the studies using structural and functional MRIs, and their respective ML methods, is given in Table~\ref{tbl:mri}.

        \begin{table*}
            \centering
            \setlength{\tabcolsep}{10pt}
            \renewcommand{\arraystretch}{2}
            \begin{tabular}{cccc}
                \textbf{Study}                             & \textbf{Year} & \textbf{Task}   & \textbf{ML Technique}    \\
                \hline 
                Gould~\etal~\cite{gould2014multivariate}           & 2014 & Multi-class Classification & SVM              \\
                \hline
                Bleich-Cohen~\etal~\cite{bleich-cohen_machine_2014}& 2014 & Multi-class Classification & SVM              \\
                \hline
                Chyzhyk~\etal~\cite{chyzhyk2015discrimination}    & 2015 & Multi-class Classification & SVM              \\
                \hline
                Tolmeijer~\etal~\cite{tolmeijer_using_2018}       & 2018 & Multi-label Regression     & Linear Regression \\
                \hline 
                Talpalaru~\etal~\cite{talpalaru_identifying_2019} & 2019 & Multi-class Classification & \begin{tabular}[c]{@{}c@{}}SVM \\ Random Forest \\ Hierarchical Clustering\end{tabular} \\
                \hline 
                Honnorat~\etal~\cite{honnorat2019neuroanatomical} & 2019 & Clustering                 &  K-Means \\
                \hline 
                Chand~\etal~\cite{chand2020two}                   & 2020 & Clustering                 & \begin{tabular}[c]{@{}c@{}}SVM \\ K-Means \end{tabular} \\
                \hline                                                                     
            \end{tabular}
            
            \caption{Summary of works relating to sub-categories of schizophrenia using MRI data and machine learning algorithms.}
            \label{tbl:mri}
        \end{table*}


        \begin{figure}[t]
            \includegraphics[width=\linewidth]{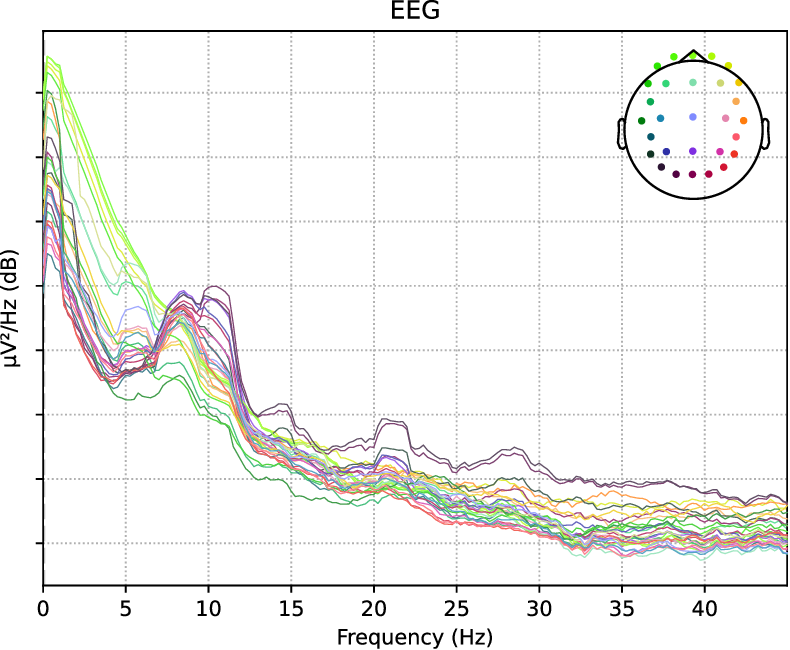}
            \caption{An example of a Power Spectral Density of EEG signals using the method of Welch~\cite{welch1967use}.}
            \label{fig:eeg}
        \end{figure}

    \subsection{EEG Input}
    An alternative method for assessing brain activity, akin to fMRI, is electroencephalography (EEG), which involves the placement of sensors on the scalp to measure electrical signals produced by the brain. Unlike fMRI, which constructs a three-dimensional image of the brain, EEG focuses on capturing and analysing the electrical activity directly.

    One significant advantage of utilising EEG is its accessibility. Unlike medical imaging techniques that require specialised equipment and trained professionals in a hospital setting, EEG can be easily obtained using simplified equipment, primarily electrodes. This flexibility allows for EEG data collection to take place anywhere, making it a more convenient option. Additionally, minimal training is required for practitioners to obtain EEG measurements effectively, making the data collection process overall more cost and time-effective.
    
    By leveraging the advantages of EEG, researchers can investigate brain activity in individuals with schizophrenia, providing valuable insights into the neuro-physiological aspects of the disorder. The accessibility and convenience of EEG make it a practical tool for studying brain dynamics and identifying potential signals associated with schizophrenia-related symptoms. A visual representation of processed EEG features and their respective nodes can be seen in Fig.~\ref{fig:eeg}.

    As in the medical imaging case, several studies~\cite{nikhil_chandran_eeg-based_2021, jahmunah_automated_2019, aslan_automatic_2020, oh_deep_2019, zhang_eeg_2019} address the problem as a binary classification between patients with schizophrenia and healthy controls. 

    Fewer works~\cite{chu_individual_2018, tikka_artificial_2020, kim_eeg_2020} attempted a more fine-grained classification of schizophrenia. Chu~\etal~\cite{chu_individual_2018} used a CNN backbone with a Random Forest classification head for a three-class classification between healthy controls, long-term patients, and first-time episodes of 140 subjects. The overall accuracy achieved is over 90\% on the three-class classifications. While the problem did not address the individual symptoms, the separation between first-time episodes and long-term patients could be an important separation for patients who have not been diagnosed timely or appropriately.

    Tikka~\etal~\cite{tikka_artificial_2020} proposed a more fine-grained classification approach that is also aligned with the PANSS~\cite{kay_positive_1987} symptom scale. Specifically, the subjects were split into healthy control, patients who scored highly on the positive symptoms, and patients who scored highly on the negative symptoms. Using hand-crafted features based on prior knowledge around areas of interest in the brain, the authors trained an SVM classifier and achieved an accuracy of 79\% and 89\% for the positive and negative classes, respectively. Similarly, Kim~\etal~\cite{kim_eeg_2020} used a simple linear classifier for binary classification between healthy controls and patients with schizophrenia using hand-crafted features. However, the authors also trained a set of classifiers that classified high-low severity for the positive, negative, and cognitive scales of the PANSS~\cite{kay_positive_1987} symptom scale.
    Such a distinction within the patient group is important, as it can affect the course of treatment, it does not however reflect the full spectrum of the condition.

        \begin{table}[h]
            \centering
            \setlength{\tabcolsep}{2pt}
            \renewcommand{\arraystretch}{3}
            \begin{tabular}{cccc}
                \textbf{Study}                             & \textbf{Year} & \textbf{Task}   & \textbf{ML Technique}    \\
                \hline 
                Chu~\etal~\cite{chu_individual_2018}     & 2018 & Multi-class Classification & \begin{tabular}[c]{@{}c@{}}CNN \\ Random Forest\end{tabular} \\
                \hline
                Tikka~\etal~\cite{tikka_artificial_2020} & 2020 & Multi-class Classification      & SVM                                                          \\
                \hline
                Kim~\etal~\cite{kim_eeg_2020}            & 2020 & Multi-class Classification     & Logistic Regression     \\
                \hline
            \end{tabular}
            \caption{Summary of work relating to subcategories of schizophrenia using EEG data and machine learning algorithms.}
            \label{tbl:eeg}
        \end{table}
    \subsection{Audio-Visual Input}
    Currently, mental health practitioners primarily rely on clinical interviews following a structured framework outlined in DSM-V~\cite{dsmd_v} to assess individuals with schizophrenia. As such, leveraging audio-visual recordings of patients for diagnosis and symptom estimation presents a more intuitive approach that closely resembles real-world conditions, than medical imaging or bio-signals.

    In clinical practice, schizophrenia manifests itself in various aspects of a patient's behaviour, encompassing facial expressions, vocal patterns, and overall demeanour. Mental health practitioners directly gauge these behavioural symptoms as an estimate of the individual's illness state and progression. Given that symptom severity can be quantified by discrete values, researchers have approached this problem as either multi-label multi-class classification or multi-label regression tasks. By employing these methodologies, it becomes feasible to capture the variations in symptom severity and provide a more detailed assessment of the patient's condition.

\begin{figure}[h]
                \includegraphics[width=\linewidth]{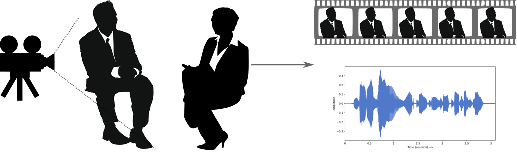}
                \caption{Flow diagram of Audio-Visual feature extraction.}
                \label{fig:av}
        \end{figure}

    A flow diagram for the extraction of Audio-Visual input is shown in Fig.~\ref{fig:av}.

    Tahir~\etal~\cite{tahir_non-verbal_2016, tahir_non-verbal_2019} addressed the problem of symptom severity estimation on the PANSS~\cite{kay_positive_1987} symptom scale, as both a classification and a regression task using SVM. More specifically, with the use of hand-crafted features of non-verbal cues associated with conversations (e.g., interruption, natural turn, etc), the authors trained an SVM and a Support Vector Regression (SVR) for the classification and regression tasks, respectively, with the latter achieving higher accuracy in the range of 53\%-80\%.

    As para-linguistic features have been proven to be crucial in estimating affect~\cite{Burkhardt_maskingspeech_2023, ren_fast_2023}, several works~\cite{chakraborty_prediction_2018, espinola_vocal_2021, boer_acoustic_2023} use low-level descriptors (lld) from audio recordings of patients. Chakraborty~\etal~\cite{chakraborty_prediction_2018}, used lld from clinical interviews and Principle Component Analysis (PCA) to reduce their dimensions. The authors trained several binary classifiers for high-low classification of each symptom on NSA16~\cite{nsa16}, achieving accuracy in the range of 61\% to 84\%. Similarly, Boer~\etal~\cite{boer_acoustic_2023} extracted acoustic features using the OpenSMILE~\cite{eyben2010opensmile} toolkit and trained a set of Random Forest classifiers for a three-class classification task, i.e., healthy versus predominantly positive versus predominantly negative, achieving an accuracy of 86.2\%.

    \begin{table*}[t]
        \centering
        \setlength{\tabcolsep}{10pt}
        \renewcommand{\arraystretch}{2}
        \begin{tabular}{cccc}
       \textbf{Study}                                                                                       & \textbf{Year} & \textbf{Task}                                                               & \textbf{ML Technique}                                                     \\
        \hline
        Tron~\etal~\cite{tron_automated_2015}                                                                & 2015 & Multi-label Regression                                                      & SVM                                                              \\
        \hline
        Tahir~\etal~\cite{tahir_non-verbal_2016}                                                             & 2016 & Multi-label Regression                                                      & \begin{tabular}[c]{@{}c@{}}SVM\\SVR\end{tabular}                 \\
        \hline
        Tron~\etal~\cite{tron_facial_2016}                                                                   & 2016 & Single-label Regression                                                     & \begin{tabular}[c]{@{}c@{}}K-Means\\SVM\end{tabular}             \\
        \hline
        Vijay~\etal~\cite{vijay2016computational}                                                            & 2016 & Multi-label Regression                                                      & SVR                                                              \\
        \hline
        Chakraborty~\etal~\cite{chakraborty_prediction_2018}                                                 & 2018 & \begin{tabular}[c]{@{}l@{}}Multi-label \\Binary Classification\end{tabular} & \begin{tabular}[c]{@{}c@{}}PCA\\Logistic Regression\end{tabular} \\
        \hline
        Tahir~\etal~\cite{tahir_non-verbal_2019}                                                             & 2019 & Multi-label Regression                                                      & \begin{tabular}[c]{@{}c@{}}SVM\\SVR\end{tabular}                 \\
        \hline
        Barzilay~\etal~\cite{Barzilay_2019_predicting}                                                       & 2019 & Multi-label Classification                                                  & SVM                                                              \\
        \hline
        Bishay~\etal~\cite{bishay_schinet_2021}                                                              & 2019 & Multi-label Regression                                                      & \begin{tabular}[c]{@{}c@{}}GMM\\Neural Network\end{tabular}      \\
        \hline
        Bishay~\etal~\cite{bishay2019can}                                                                    & 2019 & \begin{tabular}[c]{@{}l@{}}Multi-label \\Binary Classification\end{tabular} & \begin{tabular}[c]{@{}c@{}}CNN\\RNN\end{tabular}                 \\
        \hline
        Abbas~\etal~\cite{abbas_computer_2021}                                                               & 2021 & \begin{tabular}[c]{@{}l@{}}Multi-label Regression\end{tabular}               & \begin{tabular}[c]{@{}c@{}}Linear Regression\end{tabular}        \\
        \hline
        Foteinopoulou \& Patras~\cite{foteinopoulou_learning_2022}                                          & 2022 & Multi-label Regression                                                      & \begin{tabular}[c]{@{}c@{}}CNN\\Transformers\end{tabular}       \\ 
        \hline
        Boer~\etal~\cite{boer_acoustic_2023}     & 2023 & Multi-class Classification & Random Forest \\
        \hline
        \end{tabular}
        \caption{Summary of work relating to the estimation of symptoms of schizophrenia using Audio-Visual data and machine learning algorithms.}
        \label{tbl:av}
    \end{table*}

    Similar to para-linguistic audio features, certain behavioural symptoms are manifested in the subjects' facial expressions and mannerisms. Tron~\etal~\cite{tron_automated_2015, tron_facial_2016} recorded 34 schizophrenia patients and healthy controls during a clinical interview. From the video recordings, 23 Action Units (AU)~\cite{facs} and their respective intensities were extracted for each frame. In~\cite{tron_automated_2015}, hand-crafted features, such as activation ratio and intensity, were used as descriptors of the patients' facial behaviour. Using a leave-one-subject-out training/evaluation scheme, an SVM classifier was trained on binary schizophrenia detection; in addition, the authors trained a ridge regression on symptom intensity, within the patient pool of the collected dataset. The method was evaluated on all negative symptoms, three positive and two general cognitive symptoms from the PANSS~\cite{kay_positive_1987} scale, achieving a Pearson's Correlation Coefficient (PCC) of up to 53\%. Similarly, in a subsequent study~\cite{tron_facial_2016}, a $k$-means clustering algorithm was used first to assign each frame to a centroid, with the cluster centres representing facial expression prototypes. Four hand-crafted features were extracted from the number of prototypes present in each video, and the same classification and regression method were used, the latter for a single symptom achieving a PCC of 43\% on the test data. 

    The use of AU and SVR is also adopted by Vijay~\etal~\cite{vijay2016computational}; similarly to Tron~\etal~\cite{tron_automated_2015}, the authors extracted AU from the whole recording session of a patient and constructed handcrafted features related to the AU prevalence and intensity. A series of SVRs were trained using leave-one-patient-out cross-validation. Since~\cite{vijay2016computational} is an exploratory work, the predictive capabilities of the proposed model differ for each symptom, with some achieving PCC of up to 70\% (particularly for symptoms like Blunted Affect that are related to AU by definition). However, several symptoms failed to converge, particularly on the general cognitive scale.
    
    Bishay~\etal~\cite{bishay_schinet_2021} continued using AU as inputs to estimate symptom severity, as in~\cite{tron_automated_2015, tron_facial_2016, vijay2016computational}. More specifically, the authors took a staged approach, first training multiple VGG16~\cite{simonyan2014very} on the detection of individual AUs. Contrary to previous works~\cite{tron_automated_2015, tron_differentiating_2016, tron_facial_2016} that used hand-crafted features from frame level AUs,~\cite{bishay_schinet_2021} used a Gaussian Mixture Model (GMM) followed by a Fisher Vector transformation to standardise the input to a fixed-length vector. The use of GMM and Fisher Vector transformation is streamlining and automating the process further, however, there is less control in feature selection and engineering. Finally, two Fully Connected (FC) layers were then used for the regression task, the first one estimating individual symptom intensities for three symptoms on the PANSS~\cite{kay_positive_1987} negative scale or all expressive symptoms of the CAINS~\cite{forbes2010initial}, and the second estimating the total negative score using the individual symptoms as input. The method achieved a PCC of up to 42\%. Taking~\cite{bishay_schinet_2021} a step further and estimating the response of patients to treatment for individual symptoms, Bishay~\etal~\cite{bishay2019can} used stacked RNN networks to first learn a representation for each video and then, using the global (i.e., patient) representations, predicted treatment outcome achieving accuracy up to 71\% for individual symptoms on the negative scale.
    
    Foteinopoulou and Patras~\cite{foteinopoulou_learning_2022}, used a Deep Neural Network architecture on the same tasks as Bishay~\etal~\cite{bishay_schinet_2021}. Specifically, they incorporated a CNN backbone network pre-trained on the task of facial expression recognition in order to extract frame-level features. Then, they introduced a Transformer~\cite{vaswani_attention_2017}-based network that learns temporal relationships on different granularities for symptom estimation. As Transformers are prone to overfitting given limited training data and mental health datasets are typically very small, the authors proposed to address this using a novel regularisation approach. Overall, they achieved PCC up to 77\%, which is comparable to human experts as reported in~\cite{tron_automated_2015, bishay_schinet_2021}.

    Finally, as a proof-of-concept, Barzilay~\etal~\cite{Barzilay_2019_predicting} extracted features related to the subject's facial expression per frame and in the whole video. These were then used to train a set of SVM models for affect sub-types in patients with schizophrenia, one for each of the five annotators. By highlighting the high disagreement between the human annotators, this study underlines the need for automated and consistent symptom assessment and diagnosis. Furthermore, as the classification process, in this case, is (in practice) conducted using personalised models for each annotator, the overall accuracy seems to depend on the annotator; however, the method achieved accuracy up to 90\%. Similarly, as a proof-of-concept for the use of wearable technology in symptom assessment, Abbas~\etal~\cite{abbas_computer_2021} measured the head movement of subjects from smartphone front cameras. A logistic regression was used for binary classification between patients and healthy controls. A linear regression was trained for symptom severity estimation, showing a negative relationship between head movement and high-symptom severity, particularly for negative symptoms as would be expected based on the symptom definition. The results of the regression were shown to be statistically significant, with a reported p-value below 0.05.

    A summary of the works related to symptom estimation of schizophrenia using audio-visual data from clinical interviews and the machine learning methodologies used is given in Table~\ref{tbl:av}.
    
\subsection{Data collection and Annotation Methods}
\label{sec:datacollection}

 \begin{table*}[]
    \centering
    \renewcommand{\arraystretch}{3}
    \begin{tabular}{llllc}
    \textbf{Dataset} & \textbf{Annotation Availability}                     & \textbf{Data Type} & \textbf{Study}                                                                                                                                                   & \textbf{\# Samples} \\ \hline

    Proprietary                        & Coarse Categories                                                      & Medical Imaging                      & Gould~\etal~\cite{gould2014multivariate}                                                                                                                                        & 586                                    \\ \hline
    Proprietary                        & Coarse Categories                                                      & Medical Imaging                      & Bleich-Cohen~\etal~\cite{bleich-cohen_machine_2014}                                                                                                                                        & 53                                    \\ \hline
    Proprietary                        & Coarse Categories                                                      & Medical Imaging                      & Chand~\etal~\cite{chand2020two}                                                                                                                                        & 1,200                                    \\ \hline
    Proprietary                        & Fine-grained Symptoms                                                  & Medical Imaging                      & Chzhyk~\etal~\cite{chyzhyk2015discrimination}                                                                                                                                        & 68                                    \\ \hline
    Proprietary                        & Fine-grained Symptoms                                                  & Medical Imaging                      & Honnorat~\etal~\cite{honnorat2019neuroanatomical}                                                                                                                                        & 336                                    \\ \hline
    Proprietary                        & Fine-grained Symptoms                                                  & Medical Imaging                      & Tolmeijer~\etal~\cite{tolmeijer_using_2018}                                                                                                                                        & 38                                    \\ \hline
    SchizConnect~\cite{SchizConnect}     & \begin{tabular}[c]{@{}l@{}}Binary\\ Fine-grained Symptoms\end{tabular} & Medical Imaging                      & Talpalaru~\etal~\cite{talpalaru_identifying_2019}                                                   & 1,392                                 \\ \hline
    Proprietary                        & Coarse Categories                                                      & EEG                                  & Chu~\etal~\cite{chu_individual_2018}                                                                                                                                               & 120                                   \\ \hline
    Proprietary                        & Coarse Categories                                                      & EEG                                  & Tikka~\etal~\cite{tikka_artificial_2020}                                                                                                                                           & 28                                    \\ \hline
    Proprietary                        & Fine-grained Symptoms                                                  & EEG                                  & Kim~\etal~\cite{kim_eeg_2020}                                                                                                                                                      & 119                                   \\ \hline
    Proprietary                        & Fine-grained Symptoms                                                  & Audio-visual                          & \begin{tabular}[c]{@{}l@{}}Tron~\etal~\cite{tron_automated_2015}\\ Tron~\etal~\cite{tron_facial_2016}\end{tabular}                                                                 & 67                                    \\ \hline
    Proprietary                        & Fine-grained Symptoms                                                  & Audio-visual                          & Tahir~\etal~\cite{tahir_non-verbal_2016}                                                                                                                                           & 15                                    \\ \hline
    Proprietary                        & Fine-grained Symptoms                                                  & Audio-visual                          & Tahir~\etal~\cite{tahir_non-verbal_2019}                                                                                                                                           & 80                                    \\ \hline
    Proprietary                        & Fine-grained Symptoms                                                  & Audio-visual                          & Vijay~\etal~\cite{vijay2016computational}                                                                                                                                          & 18                                    \\ \hline
    Proprietary                        & Fine-grained Symptoms                                                  & Audio-visual                          & Chakraborty~\etal~\cite{chakraborty_prediction_2018}                                                                                                                               & 78                                    \\\hline
    Proprietary                        & Coarse Categories                                                      & Audio-visual                          & Barzilay~\etal~\cite{Barzilay_2019_predicting}                                                                                                                                     & 25                                    \\ \hline
    Proprietary                        & Fine-grained Symptoms                                                  & Audio-visual                          & Abbas~\etal~\cite{abbas_computer_2021}                                                                                                                                             & 27                                    \\ \hline
    NESS~\cite{ness}                   & Fine-grained Symptoms                                                  & Audio-visual                          & \begin{tabular}[c]{@{}l@{}}Bishay~\etal~\cite{bishay_schinet_2021}\\ Bishay~\etal~\cite{bishay2019can}\\ Foteinopoulou  \&  Patras~\cite{foteinopoulou_learning_2022}\end{tabular} & 110   \\
    \hline
    Proprietary                        & Coarse Categories                                                      & Audio                          & Boer~\etal~\cite{boer_acoustic_2023}                                                                                                                                     & 284                                    \\
    \end{tabular}
    \caption{Summary of Datasets used in the selected studies.}
    \label{tbl:datasets}
    \end{table*}

    In this section, we discuss the datasets used in the works included in the survey, as well as the data collection and annotation methods. A detailed view of the datasets used can be seen in Table~\ref{tbl:datasets}.
    An initial observation can be made on the number of data samples reported in each study and dataset; of the twenty-two studies and nineteen datasets included in this survey, eight or 42\% have more than 100 samples~\cite{gould2014multivariate, chand2020two, honnorat2019neuroanatomical, talpalaru_identifying_2019, chu_individual_2018, kim_eeg_2020, ness, boer_acoustic_2023} and only two or approximately 1\% have more than 1000 samples included in their studies. This is an important consideration when implementing methodologies as sample size can limit ML methodologies significantly. Furthermore, the statistical significance of the results is affected by degrees of freedom i.e. the number of samples minus the number of restrictions.

    \subsubsection{Longitudinal Studies}
    In the datasets used by the included studies, we identify three longitudinal datasets~\cite{tolmeijer_using_2018, ness, tahir_non-verbal_2016} i.e. datasets where multiple assessments are made over time to assess the progression of the illness or treatment outcome~\cite{caruana2015longitudinal}. Coincidentaly, these studies take an interventional approach i.e. they also measure treatment outcome for different types on non-pharmaceutical therapies. The remaining datasets take a cross-sectional approach~\cite{gould2014multivariate, bleich-cohen_machine_2014, chand2020two, chyzhyk2015discrimination, honnorat2019neuroanatomical, fmri1, fmri2,chu_individual_2018, tikka_artificial_2020, kim_eeg_2020, tron_automated_2015, tahir_non-verbal_2019, vijay2016computational, chakraborty_prediction_2018, Barzilay_2019_predicting, abbas_computer_2021, boer_acoustic_2023}.

    \subsubsection{Controlled vs In the wild}
    As both EEG and Medical Imaging datasets, require a controlled environment to obtain input data, we consider all of them as controlled studies. Of the remaining nine datasets used by the twelve audio-visual studies, two had a controlled recording approach using structured interviews~\cite{tron_automated_2015, tahir_non-verbal_2016} and the remaining seven take a semi-structured interview approach as outlined by DSM reflecting in the wild conditions~\cite{ness, vijay2016computational, chakraborty_prediction_2018, tahir_non-verbal_2019, Barzilay_2019_predicting, abbas_computer_2021, boer_acoustic_2023}.

    \subsubsection{Annotation Methods}
    The number of annotators is not clear on all datasets~\cite{gould2014multivariate, bleich-cohen_machine_2014, chand2020two, SchizConnect, chu_individual_2018, kim_eeg_2020, tahir_non-verbal_2019}, however, three datasets explicitly mention multiple annotators and ensuring high inter-annotation aggreement~\cite{tron_automated_2015, abbas_computer_2021, ness}. The remaining eight datasets~\cite{chyzhyk2015discrimination, honnorat2019neuroanatomical, tolmeijer_using_2018, tikka_artificial_2020, tahir_non-verbal_2016, vijay2016computational, chakraborty_prediction_2018, Barzilay_2019_predicting} explicitly or implicitly state one mental health expert annotating for each patient.

\section{Discussion and Open Questions}
\label{sec:discussion}

 The articles included in this review can be categorised into three main groups, regardless of the input type: 1) classifying or identifying schizophrenia subcategories, 2) estimating symptom severity, and 3) predicting treatment outcomes. Among the studies included nine aim to classify patients into sub-categories, ten concentrate on estimating symptom severity, and two explore the prediction of treatment outcomes. These findings suggest that, recently, researchers have adopted a more fine-grained approach to understanding the illness, with equal focus on coarse sub-types and fine-grained symptoms; less research has been done on predicting treatment outcomes using machine learning so far, which is somewhat expected as in most existing datasets standard treatment is prescribed to patients, to control for variations in outcome. Additionally, it appears that works using medical imaging and EEG tend to address the problem with more coarse labels; these can be either binary or using a few sub-categories as in the case of Chu~\etal~\cite{chu_individual_2018}, due to the availability of annotations. All methods using audio-visual input show a trend of addressing individual symptoms, either as a classification between high-low severity~\cite{chakraborty_prediction_2018} or a regression~\cite{bishay_schinet_2021}. Finally, we see that when it comes to fine-grained approaches, there are fewer recent works in the medical imaging and EEG streams, than using audio-visual input.

\subsection{Fine-grained Labels}
 This survey has focused on studies that attempt a more fine-grained approach to patient assessment than a simplistic binary classification between patients and healthy controls, as it more accurately describes the illness and real-life diagnostic conditions.

 However, as schizophrenia has several primary and secondary symptoms defined in various scales, there is a plethora of possible labels; studies in this survey each focus on a smaller sub-group with very little overlap.

 A more detailed view of the various target symptoms and the associated studies can be seen in Table~\ref{tab:target}. Symptoms from the PANSS~\cite{kay_positive_1987} seem to be most commonly used, however, no study makes a prediction on all individual symptoms of any scale. Additionally, while there is a similarity in the names and definitions between scales (eg. ``Blunted Affect'' in PANSS~\cite{kay_positive_1987}, ``Facial Expressions'' in CAINS~\cite{forbes2010initial} and ``Affect: Reduced modulation of intensity'' in NSA-16~\cite{nsa16}), the scale intensities are very different which does not allow for a direct transfer or a direct comparison between methods. Furthermore, not all studies address the symptoms in the same way with several studies~\cite{chakraborty_prediction_2018, tahir_non-verbal_2019, talpalaru_identifying_2019, kim_eeg_2020} transforming the ordinal labels to low-high categories.

 Such a binary approach to individual symptoms is a significant improvement on the simple patient vs. control classification and helps with potential class imbalance issues (assuming that the more extreme cases are also the most rare), however, it is still less detailed than real-life diagnostic and assessment criteria.

\subsection{Benefits and Limitations of Input Types}
\label{sec:input}
 Works using MRI and EEG input are the most dominant paradigm in ML for binary schizophrenia diagnosis, with fewer attempting a more fine-grained approach such as symptom estimation. These works have helped identify areas of interest in the brain~\cite{talpalaru_identifying_2019, oh_identifying_2020} that reveal important information regarding brain pathology and function, and corroborate previous findings. However, there are several open questions and practical constraints in the use of such input. First and foremost, the use of medical imaging and EEG signals are not the standard for diagnosis and assessment according to guidelines~\cite{dsmd_v}, with physicians being unable to diagnose the illness based on these sources~\cite{oh_identifying_2020}; in practice, this means that while there are several works achieving high classification accuracy in their respective datasets~\cite{jahmunah_automated_2019, aslan_automatic_2020, oh_deep_2019, zhang_eeg_2019}, medical imaging and EEG signals do not meet the current criteria for either diagnostic or illness assessment use, and so far remain exploratory works. 
 Furthermore, while high accuracy is achieved, the use of medical imaging for diagnosis and assessment is not practical as it involves significant time, costs and specialised personnel that can obtain the MRI. On the other hand, while EEG signals are significantly less costly and time-consuming to obtain, they are notoriously noisy given that the brain controls all active and inactive functions in one's body thus making signal disentanglement challenging. 

 The diagnosis and symptom assessment of schizophrenia by healthcare professionals is currently conducted over clinical interviews. Therefore, as previously discussed, methods that use audio-visual material from clinical interviews are most reflective of real-world diagnostic conditions. Furthermore, as such methods do not rely on specialised equipment and personnel, they can serve as a powerful tool to assist mental health practitioners in consistently diagnosing and assessing patients.
 However, as these are a limited snapshot of a patient's behaviour, they do not capture the wider context or pathology of the illness. Additionally, as the input is noisy and the audio-visual datasets used in the studies of this survey have more fine-grained labels than MRI and EEG datasets, the performance of such methods is not as high as in binary and coarse-label methods.

\subsection{Use of ML techniques}
\label{sec:ml}

 In terms of the use of ML techniques, most works are limited to `simpler' linear methodologies. 
 One main advantage of such approaches is the explainability aspect offered by examining the coefficients in linear models~\cite{talpalaru_identifying_2019}. Furthermore, as previously discussed, mental health datasets are typically small and therefore linear methodologies have less of a tendency to overfit. However, such methods often rely on hand-crafted features~\cite{talpalaru_identifying_2019, li_machine_2019, tolmeijer_using_2018, tron_automated_2015}, which often requires prior domain knowledge, particularly for medical imaging.
 
\onecolumn
    \begin{landscape}
    \begin{longtable}{ccccc}
    \textbf{\setlength{\tabcolsep}{2pt}\textbf{Target}}                                                                                                    & \textbf{\textbf{Task}}                                                     & \textbf{\textbf{ML Technique}}                                                   & \textbf{\textbf{Input Type}}                                 & \textbf{\textbf{Study}}                                                                                                                                                                                                                                                         \\ \hline
    \endfirsthead
    \multicolumn{5}{c}%
    {{\bfseries Table \thetable\ continued from previous page}} \\
    \textbf{\setlength{\tabcolsep}{2pt}\textbf{Target}}                                                                                                    & \textbf{\textbf{Task}}                                                     & \textbf{\textbf{ML Technique}}                                                   & \textbf{\textbf{Input Type}}                                 & \textbf{\textbf{Study}}                                                                                                                                                                                                                                                         \\ \hline
    \endhead
    \textbf{Target}                                                                                                                                        & \textbf{Task}                                                              & \textbf{ML Technique}                                                            & \textbf{Input Type}                                          & \textbf{Study}                                                                                                                                                                                                                                                                  \\
    \endhead\begin{tabular}[c]{@{}c@{}}Healthy Control \\ vs \\ Cognitive Deficit \\ vs \\ Cognitively Spared\end{tabular}                                 & Multi-class Classification                                                 & SVM                                                                              & MRI                                                          & Gould~\etal~\cite{gould2014multivariate}                                                                                                                                                                                                                                        \\ \hline
    \begin{tabular}[c]{@{}c@{}}Healthy Control \\ vs \\ Schizo-obsessive\\ vs \\ Schizophrenia\end{tabular}                                                & Multi-class Classification                                                 & SVM                                                                              & MRI                                                          & Bleich-Cohen~\etal~\cite{bleich-cohen_machine_2014}                                                                                                                                                                                                                             \\ \hline
    \begin{tabular}[c]{@{}c@{}}Healthy Control \\ vs \\ Schizophrenia Auditory Hallucinators\\ vs \\ Schizophrenia Non-Auditory Hallucinators\end{tabular} & Multi-class Classification                                                 & SVM                                                                              & MRI                                                          & Chyzhyk~\etal~\cite{chyzhyk2015discrimination}                                                                                                                                                                                                                                  \\ \hline
    Change in Positive psychotic symptoms$^a$                                                                                                              & Regression                                                     & Linear Regression                                                                & fMRI                                                         & Tolmeijer~\etal~\cite{tolmeijer_using_2018}                                                                                                                                                                                                                                     \\ \hline
    Change in Depressive symptoms$^b$                                                                                                                      & Regression                                                     & Linear Regression                                                                & fMRI                                                         & Tolmeijer~\etal~\cite{tolmeijer_using_2018}                                                                                                                                                                                                                                     \\ \hline
    \begin{tabular}[c]{@{}c@{}}Healthy Control\\ vs\\ High Symptom Burden\\ vs\\ Predominantly Positive Symptoms\\ vs \\ Mild Symptom Burden\end{tabular}$^a$& Multi-class Classification                                                 & \begin{tabular}[c]{@{}c@{}}SVM\\ Random Forest\end{tabular}                      & MRI                                                          & Talpalaru~\etal~\cite{talpalaru_identifying_2019}                                                                                                                                                                                                                               \\ \hline
    \begin{tabular}[c]{@{}c@{}}Healthy Control \\ vs \\ Fist Episode Patients\\ vs \\ High Risk Individuals\end{tabular}                                   & Multi-class Classification                                                 & \begin{tabular}[c]{@{}c@{}}CNN\\ Random Forest\end{tabular}                      & EEG                                                          & Chu~\etal~\cite{chu_individual_2018}                                                                                                                                                                                                                                            \\ \hline
    \begin{tabular}[c]{@{}c@{}}Healthy Control \\ vs\\ Scizophrenia\end{tabular}                                                                           & Binary Classification                                                      & \begin{tabular}[c]{@{}c@{}}SVM\\ Logistic Regression\end{tabular}                & \begin{tabular}[c]{@{}c@{}}EEG\\ Visual\\ Audio\end{tabular} & \begin{tabular}[c]{@{}c@{}}Tikka~\etal~\cite{tikka_artificial_2020}\\ Kim~\etal~\cite{kim_eeg_2020}\\ Tron~\etal~\cite{tron_automated_2015}\\ Tron~\etal~\cite{tron_facial_2016}\\ Tahir~\etal~\cite{tahir_non-verbal_2016}\\ Boer~\etal~\cite{boer_acoustic_2023}\end{tabular} \\ \hline
    \begin{tabular}[c]{@{}c@{}}Predominantly Positive Symptoms\\ vs\\ Predominantly Negative Symptoms\end{tabular}$^a$                                     & Binary Classification                                                      & SVM                                                                              & \begin{tabular}[c]{@{}c@{}}EEG\\ Audio\end{tabular}          & \begin{tabular}[c]{@{}c@{}}Tikka~\etal~\cite{tikka_artificial_2020}\\ Boer~\etal~\cite{boer_acoustic_2023}\end{tabular}                                                                                                                                                         \\ \hline
    \begin{tabular}[c]{@{}c@{}}High Positive\\ vs\\ Low Positive\\ vs\\ High Negative\\ vs\\ Low Negative\end{tabular}$^a$                                 & Multi-class Classification                                                 & Logistic Regression                                                              & EEG                                                          & Kim~\etal~\cite{kim_eeg_2020}                                                                                                                                                                                                                                                   \\ \hline
    N1. Blunted Affect$^a$                                                                                                                                 & Regression                                                                 & \begin{tabular}[c]{@{}c@{}}Linear Regression\\ Neural Network\end{tabular}       & Visual                                                       & \begin{tabular}[c]{@{}c@{}}Tron~\etal~\cite{tron_automated_2015}\\ Tron~\etal~\cite{tron_facial_2016}\\ Abbas~\etal~\cite{abbas_computer_2021}\\ Bishay~\etal~\cite{bishay_schinet_2021}\\ Foteinopoulou \& Patras~\cite{foteinopoulou_learning_2022}\end{tabular}              \\ \hline
    N2. Emotional withdrawal$^a$                                                                                                                           & Regression                                                                 & Linear Regression                                                                & Visual                                                       & \begin{tabular}[c]{@{}c@{}}Tron~\etal~\cite{tron_automated_2015}\\  Abbas~\etal~\cite{abbas_computer_2021}\end{tabular}                                                                                                                                                         \\ \hline
    N3. Poor Rapport$^a$                                                                                                                                   & Regression                                                                 & Linear Regression                                                                & Visual                                                       & \begin{tabular}[c]{@{}c@{}}Tron~\etal~\cite{tron_automated_2015}\\ Abbas~\etal~\cite{abbas_computer_2021}\\ Bishay~\etal~\cite{bishay_schinet_2021}\\ Foteinopoulou \& Patras~\cite{foteinopoulou_learning_2022}\end{tabular}                                                   \\ \hline
    N4. Social Withdrawal$^a$                                                                                                                              & Regression                                                                 & Linear Regression                                                                & Visual                                                       & Tron~\etal~\cite{tron_automated_2015}                                                                                                                                                                                                                                           \\ \hline
    N5. Difficulty in Abstract Thinking$^a$                                                                                                                & Regression                                                                 & Linear Regression                                                                & Visual                                                       & Tron~\etal~\cite{tron_automated_2015}                                                                                                                                                                                                                                           \\ \hline
    N6. Lack of Spontaneity$^a$                                                                                                                            & Regression                                                                 & Linear Regression                                                                & Visual                                                       & \begin{tabular}[c]{@{}c@{}}Tron~\etal~\cite{tron_automated_2015}\\ Bishay~\etal~\cite{bishay_schinet_2021}\\ Foteinopoulou \& Patras~\cite{foteinopoulou_learning_2022}\end{tabular}                                                                                            \\ \hline
    N7. Stereotyped Thinking$^a$                                                                                                                           & Regression                                                                 & Linear Regression                                                                & Visual                                                       & Tron~\etal~\cite{tron_automated_2015}                                                                                                                                                                                                                                           \\ \hline
    P1. Delusions$^a$                                                                                                                                      & Regression                                                                 & Linear Regression                                                                & Visual                                                       & Tron~\etal~\cite{tron_automated_2015}                                                                                                                                                                                                                                           \\ \hline
    P2. Conceptual Disorganization$^a$                                                                                                                     & Regression                                                                 & Linear Regression                                                                & Visual                                                       & Tron~\etal~\cite{tron_automated_2015}                                                                                                                                                                                                                                           \\ \hline
    P5. Hallucinations$^a$                                                                                                                                 & Regression                                                                 & Linear Regression                                                                & Visual                                                       & Tron~\etal~\cite{tron_automated_2015}                                                                                                                                                                                                                                           \\ \hline
    G11. Motor Retardation$^a$                                                                                                                             & Regression                                                                 & Linear Regression                                                                & Visual                                                       & \begin{tabular}[c]{@{}c@{}}Tron~\etal~\cite{tron_automated_2015}\\ Abbas~\etal~\cite{abbas_computer_2021}\end{tabular}                                                                                                                                                          \\ \hline
    G7. Poor Attention$^a$                                                                                                                                 & Regression                                                                 & Linear Regression                                                                & Visual                                                       & \begin{tabular}[c]{@{}c@{}}Tron~\etal~\cite{tron_automated_2015}\\ Abbas~\etal~\cite{abbas_computer_2021}\end{tabular}                                                                                                                                                          \\ \hline
    G8. Uncooperativeness$^a$                                                                                                                              & Regression                                                                 & Linear Regression                                                                & Visual                                                       & Abbas~\etal~\cite{abbas_computer_2021}                                                                                                                                                                                                                                          \\ \hline
    G12. Lack of judgement \& insight$^a$                                                                                                                  & Regression                                                                 & Linear Regression                                                                & Visual                                                       & Abbas~\etal~\cite{abbas_computer_2021}                                                                                                                                                                                                                                          \\ \hline
    Total Positive$^a$                                                                                                                                     & Regression                                                                 & Linear Regression                                                                & Visual                                                       & \begin{tabular}[c]{@{}c@{}}Vijay~\etal~\cite{vijay2016computational}\\ Abbas~\etal~\cite{abbas_computer_2021}\end{tabular}                                                                                                                                                      \\ \hline
    Total Negative$^a$                                                                                                                                     & Regression                                                                 & \begin{tabular}[c]{@{}c@{}}SVR\\ Linear Regression\\ Neural Network\end{tabular} & Visual                                                       & \begin{tabular}[c]{@{}c@{}}Bishay~\etal~\cite{bishay_schinet_2021}\\ Vijay~\etal~\cite{vijay2016computational}\\ Abbas~\etal~\cite{abbas_computer_2021}\\Foteinopoulou \& Patras~\cite{foteinopoulou_learning_2022}\end{tabular}                                                \\ \hline
    Total General Phychopathology$^a$                                                                                                                      & Regression                                                                 & Linear Regression                                                                & Visual                                                       & Abbas~\etal~\cite{abbas_computer_2021}                                                                                                                                                                                                                                          \\ \hline
    Facial Expression$^c$                                                                                                                                  & Regression                                                                 & Neural Network                                                                   & Visual                                                       & \begin{tabular}[c]{@{}c@{}}Bishay~\etal~\cite{bishay_schinet_2021}\\ Foteinopoulou \& Patras~\cite{foteinopoulou_learning_2022}\end{tabular}                                                                                                                                    \\ \hline
    Vocal Expression$^c$                                                                                                                                   & Regression                                                                 & Neural Network                                                                   & Visual                                                       & \begin{tabular}[c]{@{}c@{}}Bishay~\etal~\cite{bishay_schinet_2021}\\ Foteinopoulou \& Patras~\cite{foteinopoulou_learning_2022}\end{tabular}                                                                                                                                    \\ \hline
    Expressive Gestures$^c$                                                                                                                                & Regression                                                                 & Neural Network                                                                   & Visual                                                       & \begin{tabular}[c]{@{}c@{}}Bishay~\etal~\cite{bishay_schinet_2021}\\ Foteinopoulou \& Patras~\cite{foteinopoulou_learning_2022}\end{tabular}                                                                                                                                    \\ \hline
    Quantity of Speech$^c$                                                                                                                                 & Regression                                                                 & Neural Network                                                                   & Visual                                                       & \begin{tabular}[c]{@{}c@{}}Bishay~\etal~\cite{bishay_schinet_2021}\\ Foteinopoulou \& Patras~\cite{foteinopoulou_learning_2022}\end{tabular}                                                                                                                                    \\ \hline
    EXP - Total$^c$                                                                                                                                        & Regression                                                                 & Neural Network                                                                   & Visual                                                       & \begin{tabular}[c]{@{}c@{}}Bishay~\etal~\cite{bishay_schinet_2021}\\ Foteinopoulou \& Patras~\cite{foteinopoulou_learning_2022} \end{tabular}                                                                                                                                   \\ \hline
    NSA 1. Prolonged time to respond$^d$                                                                                                                   & \begin{tabular}[c]{@{}c@{}}Binary Classification\\ Regression\end{tabular} & \begin{tabular}[c]{@{}c@{}}SVM\\ SVR\\ kNN\end{tabular}                          & Audio                                                        & \begin{tabular}[c]{@{}c@{}}Tahir~\etal~\cite{tahir_non-verbal_2016}\\ Tahir~\etal~\cite{tahir_non-verbal_2019}\\ Chakraborty~\etal~\cite{chakraborty_prediction_2018}\end{tabular}                                                                                              \\ \hline
    NSA 2. Restricted speech quantity$^d$                                                                                                                  & \begin{tabular}[c]{@{}c@{}}Binary Classification\\ Regression\end{tabular} & \begin{tabular}[c]{@{}c@{}}SVM\\ SVR\\ kNN\end{tabular}                          & Audio                                                        & \begin{tabular}[c]{@{}c@{}}Tahir~\etal~\cite{tahir_non-verbal_2016}\\ Tahir~\etal~\cite{tahir_non-verbal_2019}\\ Chakraborty~\etal~\cite{chakraborty_prediction_2018}\end{tabular}                                                                                              \\ \hline
    NSA 3. Impoverished speech content$^d$                                                                                                                 & \begin{tabular}[c]{@{}c@{}}Binary Classification\\ Regression\end{tabular} & \begin{tabular}[c]{@{}c@{}}SVM\\ SVR\\ kNN\end{tabular}                          & Audio                                                        & \begin{tabular}[c]{@{}c@{}}Tahir~\etal~\cite{tahir_non-verbal_2016}\\ Tahir~\etal~\cite{tahir_non-verbal_2019}\\ Chakraborty~\etal~\cite{chakraborty_prediction_2018}\end{tabular}                                                                                              \\ \hline
    NSA 5. Emotion: Reduced range$^d$                                                                                                                      & \begin{tabular}[c]{@{}c@{}}Binary Classification\\ Regression\end{tabular} & \begin{tabular}[c]{@{}c@{}}SVM\\ SVR\\ kNN\end{tabular}                          & Audio                                                        & \begin{tabular}[c]{@{}c@{}}Tahir~\etal~\cite{tahir_non-verbal_2016}\\ Tahir~\etal~\cite{tahir_non-verbal_2019}\\ Chakraborty~\etal~\cite{chakraborty_prediction_2018}\end{tabular}                                                                                              \\ \hline
    NSA 6. Affect: Reduced modulation of intensity$^d$                                                                                                     & \begin{tabular}[c]{@{}c@{}}Binary Classification\\ Regression\end{tabular} & \begin{tabular}[c]{@{}c@{}}SVM\\ SVR\\ kNN\end{tabular}                          & Audio                                                        & \begin{tabular}[c]{@{}c@{}}Tahir~\etal~\cite{tahir_non-verbal_2016}\\ Tahir~\etal~\cite{tahir_non-verbal_2019}\\ Chakraborty~\etal~\cite{chakraborty_prediction_2018}\end{tabular}                                                                                              \\ \hline
    NSA 8. Reduced social drive$^d$                                                                                                                        & \begin{tabular}[c]{@{}c@{}}Binary Classification\\ Regression\end{tabular} & \begin{tabular}[c]{@{}c@{}}SVM\\ SVR\end{tabular}                                & Audio                                                        & \begin{tabular}[c]{@{}c@{}}Tahir~\etal~\cite{tahir_non-verbal_2016}\\ Tahir~\etal~\cite{tahir_non-verbal_2019}\end{tabular}                                                                                                                                                     \\ \hline
    NSA 15. Reduced expressive gestures$^d$                                                                                                                & \begin{tabular}[c]{@{}c@{}}Binary Classification\\ Regression\end{tabular} & \begin{tabular}[c]{@{}c@{}}SVM\\ SVR\\ kNN\end{tabular}                          & Audio                                                        & \begin{tabular}[c]{@{}c@{}}Tahir~\etal~\cite{tahir_non-verbal_2016}\\ Tahir~\etal~\cite{tahir_non-verbal_2019}\\ Chakraborty~\etal~\cite{chakraborty_prediction_2018}\end{tabular}                                                                                              \\ \hline
    Total BPRS$^e$                                                                                                                                         & Regression                                                                 & SVR                                                                              & Video                                                        & Vijay~\etal~\cite{vijay2016computational}                                                                                                                                                                                                                                       \\ \hline
    Total MADRS$^f$                                                                                                                                        & Regression                                                                 & SVR                                                                              & Video                                                        & Vijay~\etal~\cite{vijay2016computational}y                                                                                                                                                                                                                                      \\ \hline
    Quality of affect$^g$                                                                                                                                  & Multi-class Classification                                                 & SVM                                                                              & Video                                                        & Barzilay~\etal~\cite{Barzilay_2019_predicting}                                                                                                                                                                                                                                  \\ \hline
    Range of affect$^g$                                                                                                                                    & Multi-class Classification                                                 & SVM                                                                              & Video                                                        & Barzilay~\etal~\cite{Barzilay_2019_predicting}                                                                                                                                                                                                                                  \\ \hline
    Subtype of affect$^g$                                                                                                                                  & Multi-class Classification                                                 & SVM                                                                              & Video                                                        & Barzilay~\etal~\cite{Barzilay_2019_predicting} \\                                                                                                                                           \caption{Summary of symptoms and sub-categories estimated in the selected studies.}
    \label{tab:target}\\
    \end{longtable}
    \footnotesize{
    \begin{tabular}{l|l}
    $^a$ As defined in the PANSS scale~\cite{kay_positive_1987}     &   $^f$ Total Depression Score as defined in The Montgomery-Asberg Depression Rating Scale (MADRS)~\cite{montgomery1979new}\\
    $^b$ Depressive symptoms as defined in the Beck Depression Inventory~\cite{beck1996comparison}   &   $^g$ As described in in the psychiatric MSE chapter of a classic textbook of Psychiatry~\cite{normankaplan}   \\
    $^c$ As defined in the CAINS-Expressive scale~\cite{forbes2010initial} &  \\
    $^d$ As defined in the NSA-16 scale~\cite{nsa16} &  \\
    $^e$ Total score as defined in the Brief Psychiatric Rating Scale (BPRS)~\cite{overall1962brief} &  \\
    \end{tabular}
    
    }
    \end{landscape}
\twocolumn
      Additionally, as shown in other ML domains, the use of deep learning methodologies offers superior performance in a range of tasks, particularly for multi-task approaches which are not optimal in linear methods, as they lead to ambiguous areas in the decision space. Of the methods reviewed in this paper for schizophrenia symptom estimation and to our knowledge, only four employ deep learning methods~\cite{chu_individual_2018, bishay_schinet_2021, foteinopoulou_learning_2022, bishay2019can} as it can be seen on Tables~\ref{tbl:mri}~-~\ref{tbl:av} and Fig.~\ref{fig:mlovertime}. No particular shift in methodology can be identified over time as we see in Fig.~\ref{fig:mlovertime}, with SVM/SVR being the most consistently used methodologies. The trend in methods used is observed across all three input streams, and it does seem like deep learning is under-utilised. Therefore, there is a significant opportunity for more complex deep learning approaches in ML for mental health and specifically symptom estimation.

\begin{figure*}
    \centering
    \includegraphics[width=\linewidth]{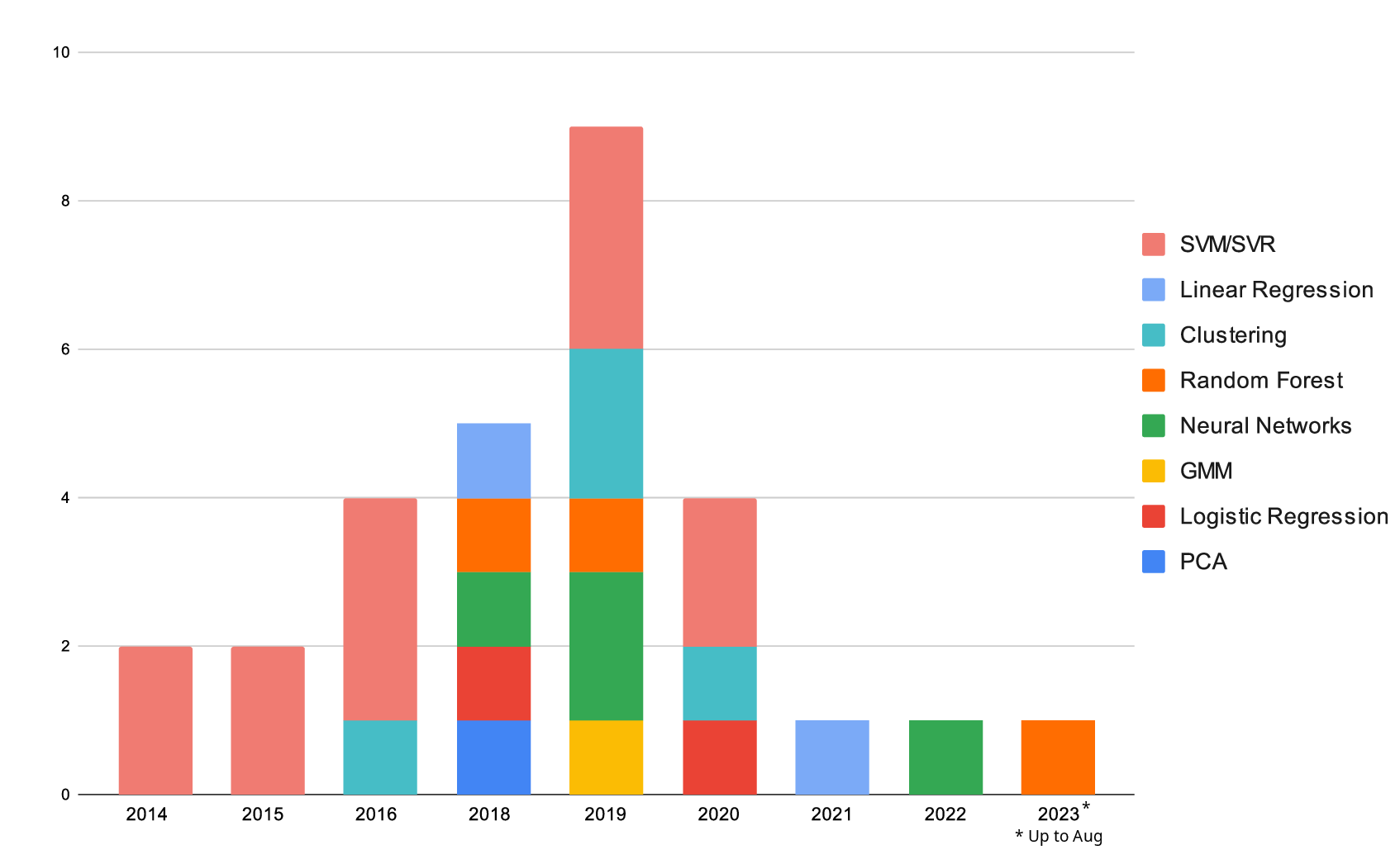}
    \caption{Number of studies per year, broken down by ML technique used. }
    \label{fig:mlovertime}
\end{figure*}

\subsection{Dataset Availability}
\label{sec:dataset}

 Significant progress has been made in recent years, in various tasks using ML methodologies, for example, Large Language Models~\cite{gpt2}; such progress has become possible due to the existence of publicly available, very large datasets. This however is not the case in most mental health tasks in general, and specifically for schizophrenia symptom estimation.
 The lack of publicly accessible benchmark datasets for evaluating and comparing different methods, is, therefore, one of the main challenges in the field of ML for mental health, particularly in relation to fine-grained schizophrenia symptom estimation. A detailed view of the datasets used in the studies included in this survey can be seen in Table~\ref{tbl:datasets} and a more detailed discussion of the collection and annotation methods used in the collected datasets can be seen in Section~\ref{sec:datacollection}.

 Although there are resources such as OpenfMRI~\cite{fmri1, fmri2} and SchizConnect~\cite{mridataset} that provide MRI inputs and annotations, these datasets are not always comprehensive and exhibit inconsistencies due to their querying nature and integration of various sources. Furthermore, while binary annotations for healthy control and schizophrenia patients are available, not all samples in these sources have fine-grained symptom annotations. Moreover, medical imaging studies often involve applying specific filters during data selection to meet patient criteria and establish associations between conditions and specific brain areas. As a result, there is significant variability in the datasets used, making it challenging to directly compare methodologies. For instance, Oh~\etal~\cite{oh_identifying_2020} and Li~\etal~\cite{li_machine_2019}, use different subsets of the COBRE~\cite{cobre} dataset obtained through SchizConnect~\cite{mridataset}. Consequently, researchers often have to collect data for their studies; this is how the majority of studies in this survey have trained and evaluated their methods. This variation makes it impossible to directly compare methodologies. Similar to MRI data, the availability of EEG datasets is also limited as most datasets are private and contain very few subjects.

 As discussed in previous sections, clinical interviews represent the most common form of diagnosing and assessing schizophrenia in real-world conditions. However, due to privacy concerns and the difficulty of anonymising such private data without losing essential information, datasets containing audio-visual inputs from clinical interviews are accessible to only a few researchers. Therefore, only a handful of works have been produced on estimating symptom severity using this type of data overall, but these works tend to approach symptom estimation using fine-grained labels. For instance, Tron~\etal~\cite{tron_automated_2015} utilise proprietary data distinct from that used by ~\cite{bishay_schinet_2021} and ~\cite{foteinopoulou_learning_2022}. As a result, significant effort is required to replicate previous works, often resulting in different performances. For example, Tron~\etal~\cite{tron_automated_2015} achieves PCC of 68\% for symptom ``N1 Blunted Affect''~\cite{kay_positive_1987}, however, the method replicated on a different set of data achieves a maximum PCC of 37\%~\cite{bishay_schinet_2021}, about half that of the original study.

 Finally, the number of samples available in all the studies included in this survey is small, particularly in comparison to recent datasets in other ML tasks where the number of samples is in the millions~\cite{schuhmann2021laion}. The low number of samples, in addition to the multi-factorial and thus noisy nature of the task, make all methods prone to overfitting and variant to domain shift thus further raising questions regarding the generalisation ability of the algorithms.


\subsection{Barriers and Future Outlook}
\label{sec:future}
 As ML for mental health is an emerging field, there are several open questions and future directions. As discussed in Section~\ref{sec:ml}, the majority of studies included in this survey use simpler linear methodologies to address the problem, with only four using deep learning approaches~\cite{ chu_individual_2018, bishay_schinet_2021, foteinopoulou_learning_2022, bishay2019can}. This is in stark contrast to the majority of works in other domains, where deep learning methodologies are the dominant paradigm. As such, there is significant potential for methodological improvements in the domain of fine-grained symptom estimation, while addressing the problems associated with mental health. Furthermore, the majority of the works use models pre-trained on different tasks to extract features, rather than using the raw signal or image in their method. Specifically, only three studies do not assume any prior knowledge during feature extraction~\cite{chakraborty_prediction_2018, espinola_vocal_2021, chu_individual_2018}, thus leading to lower omitted variable bias. Finally, the temporal relationships between features in works using audio-visual features are only utilised in two of the studies in this survey~\cite{foteinopoulou_learning_2022, chu_individual_2018}, therefore, there is significant work that can be done in the field, learning from the temporal dimension.

 In the studies included in this survey, we have identified twenty-two studies with three main input types, namely, medical imaging, EEG and audio-visual input as discussed extensively in previous sections. However, all of the works focus on a single modality to train and evaluate their method. This is counter-intuitive, as the illness itself manifests in different aspects of a patient's behaviour and physiology, which can be seen in the symptom definitions; for example, certain symptoms such as ``N1 Blunted Affect'' are by definition associated with facial expressions, while ``N6 Lack of Spontaneity and Flow in Conversation'' is more related to linguistic, para-linguistic features and dyadic interactions. Consequently, significant improvements in ML methodologies for fine-grained symptom estimation can come from future works focusing on the fusion of modalities. As the different inputs offer complementary information on the illness spectrum, multi-modal approaches are an intuitive next step towards automated, personalised assessment of schizophrenia patients, as well as mental health patients in general.
 Of course, such a research direction is significantly impacted by data availability with audio and vision modalities being easier to fuse than medical imaging and EEG where each dataset is significantly more restricted. For a more generalised multi-modal approach, a significant data collection and annotation effort needs to be made, that would allow for all complementary information to be used. A protocol for a prospective multi-modal dataset with both audio-visual and physiological data has been proposed, however, to our knowledge the data collection has not been completed and the dataset will not be publicly available due to confidentiality constraints~\cite{KONIG2022mephesto}.

 Another significant constraint in current methodologies is the context of the illness. More specifically, all datasets and methods examined in this study are either restricted to patients with schizophrenia or patients and control groups, completely disregarding other illnesses with similar symptoms or pathology. As a result, there is no indication of how these methods would perform in the real world, increasing the possibility of misdiagnosis. For example, while fMRIs can successfully distinguish between ``cognitive deficit'' and ``cognitive spared'' schizophrenia patients in Gould~\etal~\cite{gould2014multivariate}, there is no indication of whether the brain connectivity that indicates cognitive dysfunction is unique in schizophrenia patients or whether it could be manifested due to other factors that need to be ruled out first (eg. substance abuse), thus leading to misdiagnosis. Similarly, other conditions may share symptoms with schizophrenia but there is no indication of how the methods reviewed in this survey would generalise to patients with those conditions, particularly as several symptoms in schizophrenia are correlated to each other~\cite{tron_automated_2015, tron_facial_2016}, which is a form of bias in the annotations provided.
 There is therefore an opportunity for a fine-grained approach, that looks into a more general pool of subjects.

 Further to the limitations of the context, researchers in ML are faced with the continuous evolution of our understanding of the illness. As briefly discussed in previous sections, and by comparing the different editions of DSM~\cite{dsmd_v} and ICD~\cite{icd10}, we see that the definitions of symptoms or illnesses change, new symptoms are added and sub-types are dropped. This evolution in understanding is occurring for all mental health illnesses and disorders, including schizophrenia and related disorders, with every revision of diagnostic and assessment guidelines. As a result, any automated method is only a reflection of the understanding and definitions at the time of training. From a practical point of view, this translates to the need for additional data collection, annotation and training as research in mental health progresses, which can be slower and more resource-intensive than re-training human experts. As such, there is a very strong motivation for automated methodologies for fine-grained symptom estimation, using a zero-shot approach. Specifically in schizophrenia, symptom severity scales such as PANSS~\cite{kay_positive_1987} and CAINS~\cite{forbes2010initial} define not only the symptoms themselves but also the severity grades in natural language, which can act as natural language supervision for zero-shot approaches. However, these descriptions come with limitations as they can be vague and hard to conceptualise even for humans, thus requiring a very large corpus of training data to uncover the underlying pattern of behaviour. For example, the definition for Blunted Affect in PANSS~\cite{kay_positive_1987} ``Reduced range of facial expression and few expressive gestures'', however, there is no objective indication of what the normal range of expressiveness is to compare with the reduced patient expressiveness.

 Finally, regardless of the input type, there are limitations in the annotation methodology; while one of the main motivations for automated methods is the consistent diagnosis, thus avoiding the mental health professionals' bias, so far no methods address the annotator's bias and noisy labels in schizophrenia symptom estimation. In~\cite{tron_automated_2015, tahir_non-verbal_2016, tron_facial_2016, bishay_schinet_2021, foteinopoulou_learning_2022} the authors compare the PCC of the method with the agreement of the annotators, however, there is no indication of how annotators disagreement and label noise can affect the model performance. Specifically, no method addresses how the model performs when there is a large variance in the annotations and whether this variance is related to the sample noise itself. Furthermore, no indication of how the annotators' disagreement is related to the ``difficulty'' or ``rarity'' of the sample is measured. Barzilay~\etal~\cite{Barzilay_2019_predicting} and Abbas~\etal~\cite{abbas_computer_2021} train personalised models for each annotator, however, these are not consolidated into a generalised model and the variance of the annotations is not directly addressed or discussed. Label noise and annotators disagreements are addressed in other domains either with personalised models~\cite{guan_who_2018}, Gaussian processes~\cite{Long_2015_ICCV, rodrigues_gaussian_2014} or directly modelling label uncertainty as label variance~\cite{foteinopoulou_uncertainty_2021}. Consequently, for a more unbiased and consistent diagnosis and assessment, more research needs to be conducted on directly addressing annotation bias in mental health, specifically in fine-grained schizophrenia symptom severity estimation.

 The common denominator of all issues discussed in this section is the necessity for large-scale publicly available datasets with fine-grained annotations so that researchers can develop and evaluate methodologies against a common baseline. However, as we saw in Section~\ref{sec:dataset}, the majority of studies use private datasets with few samples. So far, this has been the only way for researchers to access data for most mental health tasks, due to confidentiality concerns and to avoid the identification of patients. A solution for reproducibility and fairness of evaluation, as well as availability of data to researchers, could be anonymisation via a generative methodology e.g. DeepFakes~\cite{deepfakes}. Such methods, while great at obscurification of identity, can lead to information loss (e.g. smoothing micro-expressions) or introducing noise (eg. artefacts created in the generative process), in the form of artefact creation. An alternative approach would be the publication of derived features. Such an approach, while practical, could result in omitted variable bias depending on the feature extraction method. Furthermore, as feature extraction can have a significant impact on the downstream task, a public dataset with derived latent features will need to ensure that it does not negatively affect the downstream task.

\section{Conclusion}

 In conclusion, this review paper provides a comprehensive overview of the current research on applying ML techniques for fine-grained schizophrenia sub-categories or symptom estimation. As this is an emerging field of study, the current work focuses on literature that attempts to identify sub-categories in schizophrenia or directly estimate symptom intensity. By consolidating relevant literature, this review aims to highlight the advancements in current research while identifying opportunities for future work.

 More specifically, the findings from the reviewed studies underscore the need for a more fine-grained approach to symptom estimation in schizophrenia. Particularly, works utilising medical imaging and EEG still largely address the problem with coarse labels or in a binary approach, which is insufficient to capture the nature of the illness. 

 In addition, a significant challenge hindering progress in this field is the scarcity of publicly available datasets with an adequate number of samples. More specifically, even though we identified nineteen datasets used in the studies included in this survey, only one is publicly available. In addition, over 50\% of the datasets reviewed, included less than one hundred samples and less than 1\% of the datasets reviewed included more than one thousand samples.
 The limited access to comprehensive and diverse datasets poses a barrier to the development and evaluation of ML models. 
 Furthermore, while deep learning methodologies have demonstrated remarkable capabilities in other domains, they are under-utilised in schizophrenia symptom estimation, which to some extent is a consequence of data availability. Moreover, while the reviewed studies have showcased impressive performance in symptom estimation, the generalisation ability of the proposed ML methods remains a crucial aspect to be addressed. In order to integrate ML-assisted approaches into clinical settings effectively, it is essential to demonstrate the reliability and adaptability of these models across diverse patient populations and real-world conditions, which is non-trivial to verify in the presence of private and limited data.

\section*{Acknowledgments}
The work of Niki Foteinopoulou is supported by EPSRC DTP studentship (No. EP/R513106/1) and EU H2020 AI4Media (No. 951911).

\bibliographystyle{IEEEtran}
\bibliography{ref}


\end{document}